\documentclass{revtex4}
\begin{document}
\newcommand{\dfrac}[2]{\frac{\displaystyle #1}{\displaystyle #2}}
\title{Effective potential for composite operators and for auxiliary scalar field
in a Nambu-Jona-Lasinio model
 \\}
\author{
{\bf Bang-Rong Zhou}\thanks{Electronic mailing address: zhoubr@163bj.com} }
\address{
CCAST (World Laboratory) P.O. Box 8730, Beijing 100080, China and \\
Department of physics, the Graduate School of Chinese Academy of Sciences,
Beijing 100039, China\footnote{Mailing address.}  }
\begin{abstract}
We derive the effective potentials for composite operators in a Nambu-Jona-Lasinio
(NJL) model at zero and finite temperature and show that in each case they are
equivalent to the corresponding effective potentials based on an auxiliary scalar
field. The both effective potentials could lead to the same possible spontaneous
breaking and restoration of symmetries including chiral symmetry if the momentum
cutoff in the loop integrals is large enough, and can be transformed to each other
when the Schwinger-Dyson (SD) equation of the dynamical fermion mass from the
fermion-antifermion vacuum (or thermal) condensates is used. The results also
generally indicate that two effective potentials with the same single order
parameter but rather different mathematical expressions can still be considered
physically equivalent if the SD equation corresponding to the extreme value
conditions of the two potentials have the same form.
\end{abstract}
 \maketitle

\section{Introduction}
\indent Effective potential \cite{kn:1,kn:2,kn:3,kn:4,kn:5} is a basic means to
research the vacuum of quantum field theory and spontaneous symmetry breaking. It
can be derived by means of different approaches. Consider a simple
Nambu-Jona-Lasinio (NJL) model \cite{kn:2} of four-fermion interactions
\begin{equation}
    {\cal L}(x)=\sum_{k=1}^N\bar{\psi}^k(x)i\gamma^{\mu}\partial_{\mu}\psi_k(x)+
              \frac{g}{2}\sum_{k=1}^N[\bar{\psi}^k(x)\psi_k(x)]^2,
\end{equation}
where $\psi_k(x)$  are the fermion fields with $N$ "color" components and $g$ is the
four-fermion coupling constant. The conventional approach to derive effective
potential of the model is to introduce an auxiliary scalar field $\sigma(x)$
\cite{kn:7}, since the Lagrangian (1) is equivalent to
\begin{equation}
{\cal L}_{\sigma}(x)=
\sum_{k=1}^N\bar{\psi}^k(x)i\gamma^{\mu}\partial_{\mu}\psi_k(x)
-\sigma(x)\sum_{k=1}^N\bar{\psi}^k(x)\psi_k(x) -\frac{1}{2g}\sigma^2(x),
\end{equation}
Then, In terms of a local external source $J(x)$ and the standard procedure, one
will obtain from Eq.(2) the effective action and corresponding effective potential.
The latter, in the leading order of $1/N$ expansion, can be expressed by
\cite{kn:7,kn:8}
\begin{equation}
V_{\sigma}(m_0)=\frac{m_0^2}{2g}+2N\int\frac{id^4l}{(2\pi)^4}\ln
\left(1-\frac{m_0^2}{l^2+i\varepsilon}\right),
\end{equation}
where the constant "classical field " $\sigma_c$ has been identified with the
dynamical fermion mass $m_0$. \\
\indent Alternatively, we may have another approach to derive effective potential of
the model (1), i.e. the effective action approach for composite operators presented
by Conwall, Jackiw and Tomboulis (CJT) \cite{kn:9}. In this approach, instead of
introducing an auxiliary scalar field, one considers the fermion propagator as the
order parameter of the effective potential. Since the fermion propagator $G(x,y)$ is
a bi-local function, one must put in a bi-local external source $K(x,y)$. The
effective action is the energy of the system when $G(x,y)$ is fixed. Hence in the
derivation of the effective action,  the external source $K(x,y)$ must be so
selected as to keep $G(x,y)$ to be fixed. In the final result, $G(x,y)$ will be the
exact fermion propagator, i.e. no higher order corrections to it. The CJT effective
action $\Gamma[G]$ for the models without a basic scalar field can be expressed by
\cite{kn:9,kn:10}
\begin{equation}
\Gamma[G]=-i\texttt{Tr}\ln(SG^{-1})-i\texttt{Tr}(S^{-1}G)+i\texttt{Tr}1+\Gamma_2[G],
\end{equation}
where
$$ iS^{-1}=i\gamma^{\mu}\partial_{\mu},
$$
i.e. $S$ corresponds to the propagator for free massless fermion. The $\texttt{Tr}$
is in functional sense. The first three terms in Eq.(4) are the contributions of
one-loop diagrams and $\Gamma_2[G]$ represents the contributions of two- and more-
loop vacuum graphs without fermion self-energy correction, since $G$ is the exact
fermion propagator.  In a theory of translational invariance, $G(x,y)=G(x-y)$, we
can define the effective potential $V[G]$ by
\begin{equation}
\Gamma[G]=-\Omega V[G],
\end{equation}
where $\Omega$ is the space-time volume. In the four-fermion interaction model given
by Eq.(1), we will have $V[G]=V(m_0)$, i.e. the order parameter may be replaced by
the dynamical fermion mass $m_0$. \\
\indent Several natural and interesting questions follow:  whether is the effective
potential so derived equivalent to the one from the auxiliary scalar field or not?
Whether is there any relation between the two effective potentials? What will it
mean if the answers are positive?  In this paper we will reply the above questions
through calculating CTJ potential of the model (1) and comparing it with the result
from the auxiliary scalar field approach. Besides zero temperature case, we will
also discuss finite temperature case. The discussions will be conducted in the
real-time thermal field theory \cite{kn:11,kn:12,kn:13}, and this could give us some
more insight of that how to calculate CJT effective potential in the real-time
formalism of thermal field theory, noting that conventional calculations were made
in the imaginary-time
formalism \cite{kn:14}. \\
\indent The paper is arranged as follows. In Sec.\ref{0TCJT}, we will derive the CJT
potential of the model (1) when temperature $T=0$ and fermionic chemical potential
$\mu=0$, discuss spontaneous symmetry breaking and explore the relation between the
result and Eq.(3) from the auxiliary scalar field approach. In Sec.\ref{non0TCJT},
the above discussions will be generalized to the case of finite $T$ and finite $\mu$
and in Sec.\ref{conclusion}, we give our conclusions.
\section{CJT potential at $T=\mu=0$}\label{0TCJT}
In the momentum space, we have
\begin{equation}
S(p)=i/{\not\!p}, \; G(p)=i/(\not\!{p}-m_0), \; \not\!{p}\equiv \gamma^{\mu}p_{\mu},
\end{equation}
where $G(p)$ is the exact fermion propagator when the four-fermion interactions
exist and the dynamical fermion mass $m_0$ should be a constant. When keeping only
the vacuum-vacuum diagram up to two-loop order with one four-fermion coupling vertex
in $\Gamma_2[G]$, we will obtain from Eq.(4) the CJT effective action of the model
(1) at $T=\mu=0$
\begin{eqnarray}
\Gamma[G] &=& -iN\langle[\texttt{tr}\ln
S(p)G^{-1}(p)+\texttt{tr}S^{-1}(p)G(p)-\texttt{tr}1]
       \langle p|p\rangle \rangle \nonumber\vspace{0.5cm}\\
   &&
   +\frac{g}{2}{\langle\texttt{tr}G(p)\rangle}^2(2\pi)^4\delta^4(0),
\end{eqnarray}
where $\texttt{tr}$ only represents the trace of  a spinor matrix, the denotation
$\langle\cdots\rangle$ has been used for $\int d^4p/(2\pi)^4$. By means of Eq.(6)
and the relations $\langle p|p\rangle =(2\pi)^4\delta^4(0)=\Omega$, we can write
Eq.(7) by
\begin{equation}
\Gamma[G]=-\Omega V(m_0)
\end{equation}
where, after the Wick rotation of the integral variable $p$, the effective potential
$V(m_0)$  may be expressed by
\begin{equation}
V(m_0)=-2N\left\langle\ln
\frac{\bar{p}^2+m_0^2}{\bar{p}^2}\right\rangle+4N\left\langle\frac{m_0^2}{\bar{p}^2+m_0^2}\right\rangle
-8N^2g{\left\langle\frac{m_0}{\bar{p}^2+m_0^2}\right\rangle}^2,
\end{equation}
where and afterwards, the denotation $\bar{p}$ will be understood as Euclidean 4
momentum.  By the effective potential $V(m_0)$, we may discuss spontaneous symmetry
breaking of the model (1).  A non-zero order parameter $m_0$ in the vacuum state
will mean spontaneous breaking of the discrete chiral symmetry $\chi_D$:
$\psi_k(x)\stackrel{\chi_D}{\longrightarrow}\gamma_5\psi_k(x)$ and the special
parities ${\cal P}_j$: $\psi_k(t,\cdots,x^j,\cdots) \stackrel{{\cal
P}_j}{\longrightarrow}\gamma^j\psi_k(t,\cdots,-x^j,\cdots) (j=1,2,3)$. It is
obtained from Eq.(9) that
\begin{equation}
\frac{\partial V(m_0)}{\partial
m_0}=4Nm_0\left\langle\frac{\bar{p}^2-m_0^2}{(\bar{p}^2+m_0^2)^2}\right\rangle\left(
1-4Ng\left\langle\frac{1}{\bar{p}^2+m_0^2}\right\rangle\right)
\end{equation}
with
$$\left \langle\frac{\bar{p}^2-m_0^2}{(\bar{p}^2+m_0^2)^2}\right\rangle=
\frac{m_0^2}{16\pi^2} \left[\frac{\Lambda^2}{m_0^2}
-3\ln\left(\frac{\Lambda^2}{m_0^2}+1\right)+2\frac{\Lambda^2}{\Lambda^2+m_0^2}\right],
$$
where $\Lambda$ is the 4 dimensional Euclidean momentum cutoff.  It is easy to check
that
\begin{equation}
\left\langle\frac{\bar{p}^2-m_0^2}{(\bar{p}^2+m_0^2)^2}\right\rangle > 0, \; \ {\rm
when} \ \frac{\Lambda^2}{m_0^2} > 1.82.
\end{equation}
Assuming that $\Lambda$ is large so that Eq.(11) is true, then the extreme value
condition $\partial V(m_0)/\partial m_0=0$ will be satisfied if 1) $m_0=0$ and 2)
$m_0=m_{01}$, where $m_{01}$ is determined by the gap equation
\begin{equation}
1-4Ng\left\langle\frac{1}{\bar{p}^2+m_{01}^2}\right\rangle=0.
\end{equation}
Then we may verify that
\begin{equation}
\left.\frac{\partial^2V(m_0)}{\partial
m_0^2}\right|_{m_0=0}=4N\left\langle\frac{1}{\bar{p}^2}\right\rangle
                        \left(1-4Ng\left\langle\frac{1}{\bar{p}^2}\right\rangle\right)
\end{equation}
and
\begin{equation}
\left.\frac{\partial^2V(m_0)}{\partial m_0^2}\right|_{m_0=m_{01}}=32N^2g
\left\langle\frac{\bar{p}^2-m_{01}^2}{(\bar{p}^2+m_{01}^2)^2}\right\rangle
\left\langle\frac{m_{01}^2}{(\bar{p}^2+m_{01}^2)^2}\right\rangle > 0, \; {\rm if} \
\frac{\Lambda^2}{m_{01}^2}> 1.82.
\end{equation}
Obviously, when $1-4Ng\langle 1/\bar{p}^2\rangle=1-Ng\Lambda^2/4\pi^2>0$ or
$Ng\Lambda^2/4\pi^2<1$, i.e. the four-fermion coupling $g$ is weak, $V(m_0)$ has the
only minimum point $m_0=0$ thus no spontaneous symmetry breaking occurs in this
case; Conversely, when
\begin{equation}
\frac{Ng\Lambda^2}{4\pi^2}>1,
\end{equation}
i.e. the four-fermion coupling $g$ is strong enough, $V(m_0)$ will have a maximum
point $m_0=0$ and  a minimum point $m_0=m_{01}$ which is now the non-zero solution
of Eq.(12). In this case spontaneous symmetry breaking will occur. Noting that
Eq.(15) is compatible with the condition $\Lambda^2/m_{01}^2 > 1.82$. \\
\indent We indicate that the same conclusion can be obtained by the effective
potential (3) derived from the auxiliary scalar field approach. In fact, after the
Wick rotation, Eq.(3) becomes
\begin{equation}
V_{\sigma}(m_0)=\frac{m_0^2}{2g}-2N\left\langle\ln
\frac{\bar{p}^2+m_0^2}{\bar{p}^2}\right\rangle
\end{equation}
and it further leads to
\begin{equation}
\frac{\partial V_{\sigma}(m_0)}{\partial m_0}=\frac{m_0}{g}\left(
1-4Ng\left\langle\frac{1}{\bar{p}^2+m_0^2}\right\rangle\right).
\end{equation}
It is seen by a comparison that Eq.(17) and Eq.(10) are identical except a factor
$4Ng\langle(\bar{p}^2-m_0^2)/(\bar{p}^2+m_0^2)^2\rangle$. Consequently, as far as
symmetry breaking is concerned, $V_{\sigma}(m_0)$ will lead to the same conclusion
as $V(m_0)$ when $\Lambda^2/m_0^2>1.82$. We notice that in the auxiliary scalar
field approach, the dynamical fermion mass $m_0$ comes from the vacuum expectation
value of the scalar field $\sigma(x)$ and in the CJT composite operator approach,
$m_0$ originates from the fermion-antifermion condensates
$\langle\bar{\psi}\psi\rangle$ through the relation
\begin{equation}
m_0=-\frac{g}{2}\langle\bar{\psi}\psi\rangle=
4Ng\left\langle\frac{m_0}{\bar{p}^2+m_0^2}\right\rangle.
\end{equation}
Hence if we substitute Eq.(18) into the CJT potential $V(m_0)$ in Eq.(9) and
physically this amounts to view $-\frac{g}{2}\langle\bar{\psi}\psi\rangle$
effectively as the vacuum expectation value of an auxiliary scalar field, then we
should be able to obtain $V_{\sigma}(m_0)$ from $V(m_0)$. This is in fact true.
Comparing Eq.(9) with Eq.(16) we may see that the last two terms in Eq.(9) should
correspond to the first term in the right-handed side of Eq.(16). This can be directly
verified by putting Eq.(18) into Eq.(9). \\
\indent We emphasize that the key sectors of the extreme value equations $\partial
V(m_0)/\partial m_0=0$ and $\partial V_{\sigma}(m_0)/\partial m_0=0$ are the same
and it is just Eq.(18).  This fact indicates that $V(m_0)$ and $V_{\sigma}(m_0)$ are
essentially determined by the form of the SD equation (18) and this explains that
why they are
completely equivalent despite their different expressions. \\
\section{CJT potential at finite $T$ and $\mu$}\label{non0TCJT}
The extension of the CJT effective potential (4) to finite $T$ and $\mu$ can be
expressed in the real-time formalism of thermal field theory by
\begin{eqnarray}
\Gamma_T[G]&=&-i\texttt{Tr}[\ln(S_TG_T^{-1})]^{11}-i\texttt{Tr}(S_T^{-1}G_T)^{11}
              +i\texttt{Tr}1+\Gamma_{2T}[G]  \nonumber \vspace{0.5cm}\\
  &=&-iN\langle[\texttt{tr}[\ln S_T(p)G_T^{-1}(p)]^{11}\rangle\Omega
     -iN\langle\texttt{tr}[S_T^{-1}(p)G_T(p)]^{11}\rangle\Omega
     +iN\langle\texttt{tr}1\rangle\Omega
     +\Gamma_{2T}[G],
\end{eqnarray}
where $S_T$ and $G_T$ are $2\times 2$ thermal matrix propagators, the superscript
"11" represents the 11 component of the corresponding matrix. In the momentum space,
we have \cite{kn:11}
\begin{equation}
S_T(p)=M_p\tilde{S}(p)M_p, \; G_T(p)=M_p\tilde{G}(p)M_p
\end{equation}
with the thermal transformation matrix $M_p$ defined by
\begin{equation}
M_p=\left(\matrix{\cos \theta_p & -e^{\beta \mu/2}\sin \theta_p\cr
                        e^{-\beta\mu/2}\sin \theta_p & \cos \theta_p\cr}\right), \;
\sin^2\theta_p=\frac{\theta(p^0)}{e^{\beta(p^0-\mu)}+1}
               +\frac{\theta(-p^0)}{e^{\beta(-p^0+\mu)}+1}, \; \beta=1/T
\end{equation}
and
\begin{equation}
\tilde{S}(p)=\left(\matrix{S(p)& 0 \cr
                           0   & S^*(p) \cr}\right), \;
\tilde{G}(p)=\left(\matrix{G(p)& 0 \cr
                           0   & G^*(p) \cr}\right),
\end{equation}
where
\begin{eqnarray}
  S(p)&=&i/(\not\!{p}+i\varepsilon), \; S^*(p)=-i/(\not\!{p}-i\varepsilon) \nonumber\vspace{0.3cm}\\
  G(p)&=&i/(\not\!{p}-m+i\varepsilon), \; G^*(p)=-i/(\not\!{p}-m-i\varepsilon)
\end{eqnarray}
and $m\equiv m(T,\mu)$ is the dynamical fermion mass at finite $T$ and $\mu$. Noting
that, by Eq.(20),
\begin{equation}
\ln S_T(p)G_T^{-1}(p)=\ln M_p\tilde{S}(p)\tilde{G}^{-1}(p)M_p^{-1}
                     =M_p\ln\tilde{S}(p)\tilde{G}^{-1}(p)M_p^{-1}.
\end{equation}
The last step can be checked by a formal power series expansion of the $\ln$
expression.  We may obtain from Eqs. (21)-(24) that
\begin{eqnarray}
 \texttt{tr}[\ln S_T(p)G_T^{-1}(p)]^{11}&=&\cos^2\theta_p\texttt{tr}\ln [S(p)G^{-1}(p)]
 + \sin^2\theta_p\texttt{tr}\ln [S^*(p)G^{-1*}(p)]\nonumber \\
   &=&2\cos^2\theta_p\ln \left(1-\frac{m^2}{p^2+i\varepsilon}\right)
   +2\sin^2\theta_p\ln \left(1-\frac{m^2}{p^2-i\varepsilon}\right)
\end{eqnarray}
Similarly, we may obtain
\begin{eqnarray}
\texttt{tr}[S_T^{-1}(p)G_T(p)]^{11}&=&
\texttt{tr}[M_p^{-1}\tilde{S}^{-1}(p)\tilde{G}(p)M_p]^{11}  \nonumber\\
 &=& \cos^2\theta_p\texttt{tr}[S^{-1}(p)G(p)]
 + \sin^2\theta_p\texttt{tr}[S^{-1*}(p)G^*(p)]\nonumber \\
   &=&4\left[\cos^2\theta_p\frac{p^2}{p^2-m^2+i\varepsilon}
   +\sin^2\theta_p\frac{p^2}{p^2-m^2-i\varepsilon}\right].
\end{eqnarray}
To two-loop order of the four-fermion interactions, we can calculate
\begin{eqnarray}
\Gamma_{2T}[G]&=&\frac{g}{2}(\texttt{Tr}G_T^{11})^2(2\pi)^4\delta^4(0) \nonumber\\
              &=&\frac{g}{2}N^2\left\langle\texttt{tr}G_T^{11}(p)\right\rangle^2\Omega \nonumber\\
              &=& \frac{g}{2}N^2\left\langle\texttt{tr}
                  [\cos^2\theta_pG(p)-\sin^2\theta_pG^*(p)]\right\rangle^2\Omega\nonumber\\
              &=& -8gN^2\left\langle\cos^2\theta_p\frac{m}{p^2-m^2+i\varepsilon}
   +\sin^2\theta_p\frac{m}{p^2-m^2-i\varepsilon}\right\rangle^2\Omega.
\end{eqnarray}
Substituting Eqs.(25)-(27) into Eq.(19), we will get the effective action at finite
$T$ and $\mu$
\begin{equation}
\Gamma_T[G]=-\Omega V(T,\mu,m),
\end{equation}
where
\begin{eqnarray}
 V(T,\mu, m)&=& i2N\left\langle\cos^2\theta_p\ln \left(1-\frac{m^2}{p^2+i\varepsilon}\right)
   +\sin^2\theta_p\ln \left(1-\frac{m^2}{p^2-i\varepsilon}\right)\right\rangle\nonumber \\
   &&+i4N \left\langle \cos^2\theta_p\frac{p^2}{p^2-m^2+i\varepsilon}+\sin^2\theta_p\frac{p^2}{p^2-m^2-i\varepsilon}\right\rangle \nonumber\\
   &&+8gN^2\left\langle\cos^2\theta_p\frac{m}{p^2-m^2+i\varepsilon}
   +\sin^2\theta_p\frac{m}{p^2-m^2-i\varepsilon}\right\rangle^2
\end{eqnarray}
is the effective potential at finite $T$ and $\mu$ with the order parameter $m$. By
using Eq.(29), we can discuss spontaneous symmetry breaking of the model at finite
$T$ and $\mu$. It is found out that
\begin{equation}
\frac{\partial V(T,\mu,m)}{\partial
m}=4N\left\langle\frac{i(p^2+m^2)}{(p^2-m^2+i\varepsilon)^2}\right\rangle m
\left(1-4Ng
\left\langle\frac{i}{l^2-m^2+i\varepsilon}-2\pi\delta(l^2-m^2)\sin^2\theta_l\right\rangle\right).
\end{equation}
In view of Eq.(11), we will have
$\langle{i(p^2+m^2)/(p^2-m^2+i\varepsilon)^2}\rangle>0$, if $\Lambda^2/m^2>1.82$
(after Wick rotation) is assumed. Then the extreme value condition $\partial
V(T,\mu,m)/\partial m=0$ will correspond to the equation
\begin{equation}
 m \left(1-4Ng
\left\langle\frac{i}{l^2-m^2+i\varepsilon}-2\pi\delta(l^2-m^2)\sin^2\theta_l\right\rangle\right)=0
\end{equation}
which is just the Schwinger-Dyson equation $m=-(g/2)\langle\bar{\psi}\psi\rangle_T$
obeyed by the dynamical fermion mass $m$ at finite $T$ and $\mu$, where
$\langle\bar{\psi}\psi\rangle_T$ is the thermal condensates at temperature $T$
\cite{kn:12}. The possible solutions of Eq.(31) are that 1) $m=0$ and 2) $m=m_1$,
where $m_1$ obeys the gap equation
\begin{equation}
1-4Ng
\left\langle\frac{i}{l^2-m_1^2+i\varepsilon}-2\pi\delta(l^2-m_1^2)\sin^2\theta_l\right\rangle=0.
\end{equation}
We may further find out that
\begin{equation}
\left.\frac{\partial^2V(T,\mu,m)}{\partial
m^2}\right|_{m=0}=\frac{N\Lambda^2}{4\pi^2}\left[ 1-\frac{Ng\Lambda^2}{4\pi^2}+
\frac{Ng}{2\pi^2}F_3(T,\mu,m=0) \right]
\end{equation}
with the denotations
$$
F_3(T,\mu,m)=2T^2\int_0^{\infty}\frac{dx x^2}{\sqrt{x^2+y^2}}\left[
\frac{1}{\exp(\sqrt{x^2+y^2}-r)+1}+(-r\rightarrow r) \right], \; y=\frac{m}{T},\;
r=\frac{\mu}{T}
$$
and
\begin{equation}
\left.\frac{\partial^2V(T,\mu,m)}{\partial m^2}\right|_{m=m_1}=32N^2g
\left\langle\frac{\bar{p}^2-m_1^2}{(\bar{p}^2+m_1^2)^2}\right\rangle
\left\langle\frac{m_1^2}{(\bar{p}^2+m_1^2)^2}\right\rangle > 0, \; {\rm if} \
\frac{\Lambda^2}{m_1^2}> 1.82.
\end{equation}
The gap equation (32) can be changed into
\begin{equation}
1-\frac{Ng\Lambda^2}{4\pi^2}+
\frac{Ng}{2\pi^2}\left[F_3(T,\mu,m_1)+\frac{m_1^2}{2}\ln\frac{\Lambda^2+m_1^2}{m_1^2}\right]=0
\end{equation}
and it is noted that $F_3(T,\mu,m)$ increases as $m$ goes down. We may see from
Eq.(33) that if
$$1-\frac{Ng\Lambda^2}{4\pi^2}+ \frac{Ng}{2\pi^2}F_3(T,\mu,m=0)<0,$$
then $m=0$ will be a maximum point of $V(T,\mu,m)$ and it is easy to verify that in
this case Eq.(35) could have a solution $m_1\neq 0$ which , by Eq.(34), is a minimum
point of $V(T,\mu,m)$, hence we will have spontaneous symmetry breaking at finite
$T$ and $\mu$. However, as $T$ and/or $\mu$ further increase, $F_3(T,\mu,m)$ will go
up, and by Eq.(35), $m_1$ will go down and finally to zero. As a result, we will
obtain the equation satisfied by critical temperature $T_c$ and the critical
chemical potential $\mu_c$
\begin{equation}
1-\frac{Ng\Lambda^2}{4\pi^2}+ \frac{Ng}{2\pi^2}F_3(T_c,\mu_c,m_1=0)=0.
\end{equation}
At $T_c$ and $\mu_c$, we get from Eqs. (33) and (36)
$$\left.\frac{\partial^2V(T_c,\mu_c,m)}{\partial
m^2}\right|_{m=0}=0.
$$
If $T$ and/or $\mu$ continue to go up and $F_3(T,\mu,m=0)$ will also further
increase, then we will be led to
$$\left.\frac{\partial^2V(T_c,\mu_c,m)}{\partial
m^2}\right|_{m=0}>0.
$$
This indicates that at the critical point $T_c$ and $\mu_c$,  $m=0$ will change from
being a maximum into a minimum, the order parameter $m$ also varies from nonzero to
zero. $V(T,\mu,m)$ will be left the only minimum point $m=0$ and this means that the
discrete chiral symmetry $\chi_D$ and the special parities ${\cal P}_j
 (j=1,2,3)$ which are spontaneously broken at $T=\mu=0$ and $T<T_c$ and/or $\mu<\mu_c$
will be restored. However, the above discussions do not involve the order of the
phase transition. For the latter more demonstrations
are needed \cite{kn:15}. \\
\indent Similar to the zero temperature case, the above conclusions coming from the
CJT potential $V(T,\mu.m)$ can also be reached by the effective potential
$V_{\sigma}(T,\mu,m)$ based on the auxiliary scalar field. The effective potential
of a NJL model at finite $T$ and $\mu$ based on auxiliary field was derived for the
first time in Ref.\cite{kn:16}. For the model (1), $V_{\sigma}(T,\mu,m)$ and
relevant expressions can be written by
\begin{equation}
V_{\sigma}(T,\mu,m)=\frac{m^2}{2g}+i2N\left\langle\cos^2\theta_p\ln
\left(1-\frac{m^2}{p^2+i\varepsilon}\right)
   +\sin^2\theta_p\ln \left(1-\frac{m^2}{p^2-i\varepsilon}\right)\right\rangle
\end{equation}
\begin{equation}
\frac{\partial V_{\sigma}(T,\mu,m)}{\partial m}=\frac{m}{g} \left(1-4Ng
\left\langle\frac{i}{p^2-m^2+i\varepsilon}-2\pi\delta(p^2-m^2)\sin^2\theta_p\right\rangle\right)
\end{equation}
\begin{equation}
\left.\frac{\partial^2V_{\sigma}(T,\mu,m)}{\partial
m^2}\right|_{m=0}=\frac{1}{g}\left[ 1-\frac{Ng\Lambda^2}{4\pi^2}+
\frac{Ng}{2\pi^2}F_3(T,\mu,m=0) \right]
\end{equation}
\begin{equation}
\left.\frac{\partial^2V_{\sigma}(T,\mu,m)}{\partial m^2}\right|_{m=m_1}= 8m_1^2
\left\langle\frac{-i}{(p^2-m_1^2+i\varepsilon)^2}\right\rangle
> 0,
\end{equation}
where $m_1$ is the non-zero solution of $\partial V_{\sigma}(T,\mu,m)/\partial m=0$.
Obviously Eqs.(37)-(40) will reproduce the total conclusions from the CJT potential
$V(T,\mu,m)$. This indicates that at finite $T$ and $\mu$, the two potentials are
also completely equivalent. In fact, if substituting the SD equation (31) from the
thermal condensates $\langle\bar{\psi}\psi\rangle_T$ into Eq.(29), then we will
reduce $V(T,\mu,m)$ to $V_{\sigma}(T,\mu,m)$ given by Eq.(37). This shows that the
following two approaches will lead to physically identical results: one is that
after the thermal condensates $\langle\bar{\psi}\psi\rangle_T$ are formed, to
consider them as an effective constant scalar field to generate the dynamical
fermion mass $m$, and another one is that from the beginning of the discussions, to
replace the four-fermion interactions by the effective Yukawa coupling between an
auxiliary scalar field and the fermions so as to spontaneously obtain the fermion
mass $m$. Although the CJT potential $V(T,\mu,m)$ and the auxiliary scalar field
effective potential $V_{\sigma}(T,\mu,m)$ may have different expressions, the key
sector contained in the both are the same. That is the derived Schwinger-Dyson
equation (31) from the thermal condensates $\langle\bar{\psi}\psi\rangle_T$.
\section{Conclusions}\label{conclusion}
We have proven that in a 4D NJL model, the CJT effective potential based on
composite operators and the effective potential based on an auxiliary scalar field
are completely equivalent not only at $T=\mu=0$ but also at a finite $T$ and $\mu$
if the momentum cutoff in the fermion loop integrals is large enough. The two
effective potentials may give the same possible spontaneous symmetry (including
chiral symmetry) breaking at $T=\mu=0$ and a low $T$ and $\mu$, and symmetry
restoration at a finite $T$ and $\mu$. Although the expressions of the two effective
potentials are different, the key sectors of the both are the same, it is the
derived Schwinger-Dyson equation originated from the fermion -antifermion
condensates. In particular, when the above equation is used, the two potentials can
be transformed to each other. The discussions also imply that, in general, an
effective potential, at least the one containing a single order parameter, is
essentially determined by the Schwinger-Dyson equation corresponding to the extreme
condition of the effective potential.  This is because the mathematical form of the
SD equation will determine the effective potential's extreme value points, maximums,
minimums, etc. and they are essential for research on spontaneous symmetry breaking.
Hence, corresponding to a given form of the SD equation, it seems that we are
allowed to have different expressions for the corresponding effective potential
\cite{kn:8}. \acknowledgements This work was supported by the National Natural
Science Foundation of China.


\begin{references}
\bibitem{kn:1} W. Heisenberg and H. Euler, Z. Phys. {\bf 98}, 714 (1936);
               J. Schwinger, Phys. Rev. {\bf 82}, 664 (1951).
\bibitem{kn:2} J. Goldstone, A. Salam, and J. Weinberg, Phys. Rev. {\bf 127}, 965
               (1962);
               G. Jona-Lasinio, N. Cimento {\bf 34}, 1790 (1964).
\bibitem{kn:3} S. Coleman and E. Weinberg, Phys. Rev. D{\bf 7}, 1888 (1973).
\bibitem{kn:4} R. Jackiw, Phys. Rev. D{\bf 9}, 1686 (1974).
\bibitem{kn:5} J. Iliopoulos, C. Itzykson, and A. Martin, Rev. Mod. Phys. {\bf 47},
               165 (1975).
\bibitem{kn:6} Y. Nambu and G. Jona-Lasinio, Phys. Rev. {\bf 122}, 345 (1961);
               {\bf 124}, 246 (1961).
\bibitem{kn:7} D. Ebert and H. Reinhardt, Nucl. Phys. B{\bf 271}, 188 (1986);
               B. Rosenstein, B. J. Warr, and S. H. Park, Phys. Rep. {\bf 205},
                59 (1991) and references therein;
               V. A. Miransky, {\it Dynamical symmetry breaking in quantum field
               theory} (World Scientific, Singapore, 1993).
\bibitem{kn:8} Zhou Bang-Rong, Commun. Theor. Phys. {\bf 39}, 663 (2003).
\bibitem{kn:9} J. M. Cornwall, R. Jackiw, and E. Tomboulis, Phys. Rev. D{\bf 10},
               2428 (1974).
\bibitem{kn:10} K. Higashijima, Prog. Theor. Phys. Suppl. {\bf 104}, 1 (1991).
\bibitem{kn:11} N. P. Landsman and Ch. G. van Weert, Phys. Rep. {\bf 145}, 141 (1987).
\bibitem{kn:12} Bang-Rong Zhou, Phys. Rev. D{\bf 57}, 3171 (1998); Commun. Theor.
                Phys. {\bf 32}, 425 (1999).
\bibitem{kn:13} Bang-Rong Zhou, Phys. Rev. D{\bf 59}, 065007(1999); D{\bf 62}, 105004
                (2000); D{\bf 65}, 027701 (2002); Commun. Theor. Phys. {\bf 37}, 685 (2002).
\bibitem{kn:14} O. Kiriyama, M. Maruyama, and F. Takagi, Phys. Rev. D{\bf 62}, 105008 (2000).
\bibitem{kn:15} Zhou Bang-Rong, Commun. Theor. Phys. {\bf 40}, 67 (2003); {\bf 40}, 669 (2003).
\bibitem{kn:16} H. Reinhardt and B. V. Dang, J. Phys. G{\bf 13}, 1179 (1987).
\end{references}
\end{document}